\documentclass{ws-mpla}

\newcommand{\Eqn}[1]{&\hspace{-0.2em}#1\hspace{-0.2em}&}
\def\Vec#1{\mbox{\boldmath $#1$}}

\def\be{\begin{equation}}
\def\ee{\end{equation}}
\def\bea{\begin{eqnarray}}
\def\eea{\end{eqnarray}}
\def\beq{\begin{eqnarray}}
\def\eeq{\end{eqnarray}}

\def\be{\begin{equation}}
\def\ee{\end{equation}}
\def\bea{\begin{eqnarray}}
\def\eea{\end{eqnarray}}
\def\beq{\begin{eqnarray}}
\def\eeq{\end{eqnarray}}

\begin{document}

\markboth{
Kazuharu Bamba, Chao-Qiang Geng, Shin'ichi Nojiri and Sergei D. Odintsov
}
{
Crossing of Phantom Divide in $F(R)$ Gravity
}

\catchline{}{}{}{}{}

\title{
Crossing of Phantom Divide in $F(R)$ Gravity
}

\author{\footnotesize 
KAZUHARU BAMBA
}

\address{Department of Physics, National Tsing Hua University, Hsinchu, 
Taiwan 300, R. O. C.\\
bamba@phys.nthu.edu.tw}

\author{
CHAO-QIANG GENG
}

\address{Department of Physics, National Tsing Hua University, Hsinchu, 
Taiwan 300, R. O. C.\\
geng@phys.nthu.edu.tw
}

\author{
SHIN'ICHI NOJIRI
}

\address{Department of Physics, Nagoya University, Nagoya 464-8602, 
Japan\\
nojiri@phys.nagoya-u.ac.jp
}

\author{
SERGEI D. ODINTSOV\footnote{
Also at Tomsk State Pedagogical University}} 
\address{Instituci\`{o} Catalana de Recerca i Estudis Avan\c{c}ats (ICREA)
and Institut de Ciencies de l'Espai (IEEC-CSIC),
Campus UAB, Facultat de Ciencies, Torre C5-Par-2a pl, E-08193 Bellaterra
(Barcelona), Spain\\
odintsov@aliga.ieec.uab.es
}

\maketitle

\pub{Received (Day Month Year)}{Revised (Day Month Year)}

\begin{abstract}

An explicit model of $F(R)$ gravity with realizing 
a crossing of the phantom divide is reconstructed. 
In particular, it is shown that the Big Rip singularity 
may appear in the reconstructed model of $F(R)$ gravity. 
Such a Big Rip singularity could be avoided by adding $R^2$ term 
or non-singular viable $F(R)$ theory\cite{Nojiri:2009xw} 
to the model because phantom behavior becomes transient. 

\keywords{
Modified theories of gravity; 
Dark energy; 
Cosmology.}
\end{abstract}

\ccode{PACS Nos.: 
04.50.Kd, 95.36.+x, 98.80.-k
}

\section{Introduction}	

It is observationally supported that the current expansion of the universe
is accelerating.\cite{Spergel:2003cb,Peiris:2003ff,Spergel:2006hy,Komatsu:2008hk,Perlmutter:1998np,Riess:1998cb,Astier:2005qq,Riess:2006fw} 
A number of scenarios to account for the current accelerated expansion of 
the universe have been proposed 
(for reviews, see Refs.~\refcite{Peebles:2002gy,Sahni:2005ct,Padmanabhan:2002ji,Copeland:2006wr,review,Nojiri:2008nk,N-O-2,S-F,Lobo-08,C-F}). 

Approaches to explain the current accelerated expansion of 
the universe fall into two broad categories. One is the introduction of 
some unknown matter, which is called 
``dark energy'' in the framework of general relativity.
The other is the modification of the gravitational theory, e.g., 
``$F(R)$ gravity'', where 
$F(R)$ is an arbitrary function of the scalar curvature $R$ 
(for reviews, 
see Refs.~\refcite{review,Nojiri:2008nk,N-O-2,S-F,Lobo-08,C-F}).

Recent various observational 
data\cite{Alam:2004jy,Nesseris:2006er,Wu:2006bb,J-B-P} 
imply that the effective equation of state (EoS), 
which is the ratio of the effective pressure of the universe to 
the effective energy density of it, 
may evolve from larger than $-1$ (non-phantom phase) to less than $-1$
(phantom one), namely, 
cross $-1$ (the phantom divide). 

Various investigations to realize the crossing of the phantom divide have 
been executed in the framework of general relativity: 
Scalar-tensor theories with the non-minimal gravitational coupling
between a scalar field and the scalar curvature 
or that between a scalar field and the Gauss-Bonnet 
term, 
one scalar field model with non-linear kinetic terms 
or a non-linear higher-derivative one, 
phantom coupled to dark matter with an appropriate 
coupling, 
the thermodynamical inhomogeneous dark energy model, 
multiple kinetic k-essence, 
multi-field models (two scalar fields model, 
``quintom'' consisting of phantom and canonical scalar
fields), and the description of those models through 
the Parameterized Post-Friedmann approach, 
or a classical Dirac field 
or 
string-inspired models, 
non-local gravity, 
a model in loop quantum cosmology 
and a general consideration of the crossing of the phantom 
divide 
(for a detailed review, see Ref.~\refcite{Copeland:2006wr}). 
However, explicit models of modified gravity realizing the
crossing of the phantom divide have hardly been examined, although 
there were suggestive and interesting related 
works.\cite{review,abdalla,Amendola:2007nt} 

In the present paper, we review our results in Ref.~\refcite{Bamba:2008hq} and 
reconstruct an explicit model of $F(R)$ gravity in which a crossing of the 
phantom divide can be realized by using the reconstruction method proposed in 
Refs.~\refcite{Nojiri:2006gh,Nojiri:2006be} 
(for more detailed references, see references in 
Refs.~\refcite{Bamba:2008hq,Bamba:2009ay,Bamba:2009kc,Bamba:2009vq,Bamba:2009dk}). 
It is demonstrated that the Big Rip singularity may appear in the 
reconstructed model of $F(R)$ gravity.

\section{Reconstruction of a $F(R)$ gravity model 
with realizing a crossing of the phantom divide}

\subsection{Reconstruction method}

To begin with, we briefly review the reconstruction method of modified
gravity.\cite{Nojiri:2006gh,Nojiri:2006be} 

The action of $F(R)$ gravity with general matter is given by
\begin{eqnarray}
S = \int d^4 x \sqrt{-g} \left[ \frac{F(R)}{2\kappa^2} +
{\mathcal{L}}_{\mathrm{matter}} \right]\,,
\label{eq:2.1}
\end{eqnarray}
where $g$ is the determinant of the metric tensor $g_{\mu\nu}$ and
${\mathcal{L}}_{\mathrm{matter}}$ is the matter Lagrangian.

By using proper functions $P(\phi)$ and $Q(\phi)$ 
of a scalar field $\phi$, 
the action in Eq.~(\ref{eq:2.1}) can be rewritten to the following form:
\begin{eqnarray}
S=\int d^4 x \sqrt{-g} \left\{ \frac{1}{2\kappa^2} \left[ P(\phi) R + Q(\phi)
\right] + {\mathcal{L}}_{\mathrm{matter}} \right\}\,.
\label{eq:2.2}
\end{eqnarray}
The scalar field $\phi$ may be regarded as an auxiliary scalar field because
$\phi$ has no kinetic term. 
{}From the action in Eq.~(\ref{eq:2.1}), the equation of motion of
$\phi$ is derived as 
\begin{eqnarray}
0=\frac{d P(\phi)}{d \phi} R + \frac{d Q(\phi)}{d \phi}\,,
\label{eq:2.3}
\end{eqnarray}
which may be solved with respect to $\phi$ as $\phi=\phi(R)$. 
Substituting $\phi=\phi(R)$ into the action (\ref{eq:2.2}), 
we find that the expression of $F(R)$ in the action of $F(R)$ gravity in 
Eq.~(\ref{eq:2.1}) is given by
\begin{eqnarray}
F(R) = P(\phi(R)) R + Q(\phi(R))\,.
\label{eq:2.4}
\end{eqnarray}

{}From the action in Eq.~(\ref{eq:2.2}), the gravitational field equation 
is given by 
\begin{equation}
\frac{1}{2}g_{\mu \nu} \left[ P(\phi) R + Q(\phi) \right]
-R_{\mu \nu} P(\phi) -g_{\mu \nu} \Box P(\phi) +
{\nabla}_{\mu} {\nabla}_{\nu}P(\phi) + \kappa^2
T^{(\mathrm{matter})}_{\mu \nu} = 0\,,
\label{eq:2.5}
\end{equation}
where ${\nabla}_{\mu}$ is the covariant derivative operator associated with
$g_{\mu \nu}$, $\Box \equiv g^{\mu \nu} {\nabla}_{\mu} {\nabla}_{\nu}$
is the covariant d'Alembertian for a scalar field, and
$T^{(\mathrm{matter})}_{\mu \nu}$ is the contribution to
the matter energy-momentum tensor.

We assume the flat 
Friedmann-Robertson-Walker (FRW) space-time with the metric,
\begin{eqnarray}
{ds}^2 = -{dt}^2 + a^2(t)d{\Vec{x}}^2\,,
\label{eq:2.6}
\end{eqnarray}
where $a(t)$ is the scale factor.

In the flat FRW background (\ref{eq:2.6}), the $(\mu,\nu)=(0,0)$ component and 
the trace part of the $(\mu,\nu)=(i,j)$ component of Eq.~(\ref{eq:2.5}), 
where $i$ and $j$ run from $1$ to $3$, become 
\begin{eqnarray}
-6H^2P(\phi(t)) -Q(\phi(t)) -6H \frac{dP(\phi(t))}{dt} + 2\kappa^2\rho = 0\,,
\label{eq:2.7}
\end{eqnarray}
and
\begin{eqnarray}
2\frac{d^2P(\phi(t))}{dt^2}+4H\frac{dP(\phi(t))}{dt}+
\left(4\dot{H}+6H^2 \right)P(\phi(t)) +Q(\phi(t)) + 2\kappa^2 p = 0\,,
\label{eq:2.8}
\end{eqnarray}
respectively, 
where $H=\dot{a}/a$ is the Hubble parameter and a dot denotes a time
derivative, $\dot{~}=\partial/\partial t$.
Here, $\rho$ and $p$ are the sum of the energy density and
pressure of matters with a constant
EoS parameter $w_i$, respectively, where $i$ denotes some component of
the matters.

By eliminating $Q(\phi)$ from Eqs.~(\ref{eq:2.7}) and (\ref{eq:2.8}), 
we obtain
\begin{eqnarray}
\frac{d^2P(\phi(t))}{dt^2} -H\frac{dP(\phi(t))}{dt} +2\dot{H}P(\phi(t)) +
\kappa^2 \left( \rho + p \right) = 0\,.
\label{eq:2.9}
\end{eqnarray}
We note that the scalar field $\phi$ may be taken as $\phi = t$ because 
$\phi$ can be redefined properly.

We consider that $a(t)$ is expressed as 
\begin{eqnarray}
a(t) = \bar{a} \exp \left( \tilde{g}(t) \right)\,,
\label{eq:2.10}
\end{eqnarray}
where
$\bar{a}$ is a constant and $\tilde{g}(t)$ is a proper function.
In this case, Eq.~(\ref{eq:2.9}) is reduced to
\begin{eqnarray}
&&
\frac{d^2P(\phi)}{d\phi^2} -\frac{d \tilde{g}(\phi)}{d\phi}
\frac{dP(\phi)}{d\phi} +2 \frac{d^2 \tilde{g}(\phi)}{d \phi^2}
P(\phi) \nonumber \\
&& \hspace{10mm}
{}+
\kappa^2 \sum_i \left( 1+w_i \right) \bar{\rho}_i
\bar{a}^{-3\left( 1+w_i \right)} \exp
\left[ -3\left( 1+w_i \right) \tilde{g}(\phi) \right] = 0\,,
\label{eq:2.11}
\end{eqnarray}
where $\bar{\rho}_i$ is a constant and we have used
$H= d \tilde{g}(\phi)/\left(d \phi \right)$.
Moreover, it follows from Eq.~(\ref{eq:2.7}) that $Q(\phi)$ is given by
\begin{eqnarray}
Q(\phi) \Eqn{=} -6 \left[ \frac{d \tilde{g}(\phi)}{d\phi} \right]^2 P(\phi)
-6\frac{d \tilde{g}(\phi)}{d\phi} \frac{dP(\phi)}{d\phi} \nonumber \\
&& \hspace{10mm}
{}+
2\kappa^2 \sum_i \bar{\rho}_i \bar{a}^{-3\left( 1+w_i \right)}
\exp
\left[ -3\left( 1+w_i \right) \tilde{g}(\phi) \right]\,.
\label{eq:2.12}
\end{eqnarray}
Hence, if the solution of Eq.~(\ref{eq:2.11}) with respect to 
$P(\phi)$ is obtained, we can find $Q(\phi)$.

\subsection{Explicit $F(R)$ gravity model with realizing a crossing of the 
phantom divide}

Next, we reconstruct an explicit model of $F(R)$ gravity in which 
a crossing of the phantom divide can be realized 
by using the reconstruction method explained in the preceding subsection. 

A solution of Eq.~(\ref{eq:2.11}) without matter is given by 
\begin{eqnarray}
P(\phi) \Eqn{=} e^{\tilde{g}(\phi)/2} \tilde{p}(\phi)\,, 
\label{PDF2} \\
\tilde{g}(\phi) \Eqn{=} - 10 \ln \left[ \left(\frac{\phi}{t_0}\right)^{-\gamma}
 - C \left(\frac{\phi}{t_0}\right)^{\gamma+1} \right]\,, 
\label{PDF4} \\ 
\tilde{p}(\phi) \Eqn{=} \tilde{p}_+ \phi^{\beta_+} + 
\tilde{p}_- \phi^{\beta_-}\,, 
\label{PDF6} \\
\beta_\pm \Eqn{=} \frac{1 \pm \sqrt{1 + 100 \gamma (\gamma + 1)}}{2}\,,
\label{PDF7}
\end{eqnarray}
where $\gamma$ and $C$ are positive constants, 
$t_0$ is the present time, and $\tilde{p}_\pm$ are arbitrary constants. 

{}From Eq.~(\ref{PDF4}), we see that $\tilde{g}(\phi)$ diverges at
finite $\phi$ when
\be
\label{PDF8}
\phi = t_s \equiv t_0 C^{-1/(2\gamma + 1)}\ ,
\ee
which implies that there could be the Big Rip singularity at 
$t=t_s$. 

We consider only the period $0<t<t_s$ because
$\tilde{g}(\phi)$ should be real number. 
{}From Eq.~(\ref{PDF4}), we obtain the following Hubble rate $H(t)$: 
\be
\label{PDF9}
H(t)= \frac{d \tilde{g}(\phi)}{d \phi}
= \left(\frac{10}{t_0}\right) \left[ \frac{ \gamma \left(\frac{\phi}{t_0}\right)^{-\gamma-1 }
 + (\gamma+1) C \left(\frac{\phi}{t_0}\right)^{\gamma} }{\left(\frac{\phi}{t_0}\right)^{-\gamma}
 - C \left(\frac{\phi}{t_0}\right)^{\gamma+1}}\right]\ ,
\ee
where it is taken $\phi=t$.

In the flat FRW background (\ref{eq:2.6}), even for $F(R)$ gravity 
described by the action in Eq.~(\ref{eq:2.1}), the effective energy-density 
and pressure of the universe are given by 
$\rho_\mathrm{eff} = 3H^2/\kappa^2$ and
$p_\mathrm{eff} = -\left(2\dot{H} + 3H^2 \right)/\kappa^2$, respectively.
The effective EoS 
$w_\mathrm{eff} = p_\mathrm{eff}/\rho_\mathrm{eff}$
is defined as\cite{review}
\begin{eqnarray}
w_\mathrm{eff} \equiv -1 -\frac{2\dot{H}}{3H^2}\,.
\label{eq:2.16}
\end{eqnarray}
For the case of $H(t)$ in Eq.~(\ref{PDF9}), from Eq.~(\ref{eq:2.16}) we
find that $w_\mathrm{eff}$ is expressed as
\begin{eqnarray}
w_\mathrm{eff} = -1 + U(t)\,,
\label{eq:I0-1-1}
\end{eqnarray}
where
\begin{eqnarray}
U(t) \equiv -\frac{2\dot{H}}{3H^2} =
- \frac{-\gamma + 4\gamma \left( \gamma+1 \right)
\left( \frac{t}{t_s} \right)^{2\gamma+1} +
\left( \gamma+1 \right) \left( \frac{t}{t_s}
\right)^{2\left( 2\gamma+1 \right)} }
{15 \left[ \gamma + \left( \gamma+1 \right)
\left( \frac{t}{t_s} \right)^{2\gamma+1} \right]^2}\,.
\label{eq:I0-1-2}
\end{eqnarray} 
Furthermore, 
the scalar curvature is given by $R=6\left( \dot{H} + 2H^2 \right)$. 
For the case of Eq.~(\ref{PDF9}), $R$ is described as
\begin{equation}
R = 
\frac{60
\Biggl[
\gamma \left( 20\gamma -1 \right) + 44\gamma \left( \gamma+1 \right)
\left( \frac{t}{t_s} \right)^{2\gamma+1} 
+ \left( \gamma+1 \right) \left( 20\gamma+21 \right)
\left( \frac{t}{t_s} \right)^{2\left( 2\gamma+1 \right)}
\Biggr]
}
{t^2 \left[ 1- \left( \frac{t}{t_s} \right)^{2\gamma+1} \right]^2}
\,.
\label{eq:I0-2}
\end{equation} 
In deriving Eqs.~(\ref{eq:I0-1-2}) and (\ref{eq:I0-2}), we have used
Eq.~(\ref{PDF8}).

When $t\to 0$, i.e., $t \ll t_s$, $H(t)$ behaves as
\be
\label{PDF10}
H(t) \sim \frac{10\gamma}{t}\ .
\ee
In this limit, it follows from Eq.~(\ref{eq:2.16}) that
the effective EoS parameter is given by
\be
\label{PDF11}
w_\mathrm{eff} = -1 + \frac{1}{15\gamma}\ .
\ee
This behavior is 
identical with that in the Einstein gravity with matter
whose EoS is greater than $-1$. 

On the other hand, when $t\to t_s$, we find
\be
\label{PDF12}
H(t) \sim \frac{10}{t_s - t}\ .
\ee 
In this case, the scale factor is given by
$a(t) \sim \bar{a} \left( t_s - t \right)^{-10}$.
When $t\to t_s$, therefore, $a \to \infty$, namely, the Big Rip singularity
appears.
In this limit, the effective EoS parameter is given by
\be
\label{PDF13}
w_\mathrm{eff} = - 1 - \frac{1}{15} = -\frac{16}{15}\ .
\ee 
This behavior is identical with the case in which there is a phantom matter
with its EoS being smaller than $-1$. 
As a consequence, we have reconstructed an explicit model with realizing a 
crossing of the phantom divide. 

{}From Eq.~(\ref{eq:2.16}), we see that the effective EoS 
$w_\mathrm{eff}$ becomes $-1$ when $\dot{H}=0$. 
By solving $w_\mathrm{eff} = -1$ with respect to
$t$ by using Eq.~(\ref{eq:I0-1-1}), namely, $U(t)=0$, 
we find that the effective EoS parameter crosses the phantom divide at 
$t=t_\mathrm{c}$, given by 
\begin{eqnarray}
t_\mathrm{c} = t_s \left( -2\gamma +
\sqrt{4\gamma^2 + \frac{\gamma}{\gamma+1}}
\right)^{1/\left( 2\gamma + 1 \right)}\,.
\label{eq:I1}
\end{eqnarray} 
It follows from Eq.~(\ref{eq:I0-1-2}) that when $t<t_\mathrm{c}$, $U(t)>0$
because $\gamma >0$. 
Moreover, the time derivative of $U(t)$ is given by
\begin{eqnarray}
\frac{d U(t)}{dt} =
-\frac{2\gamma \left( \gamma+1 \right) \left( 2\gamma+1 \right)^2}
{15\left[ \gamma + \left( \gamma+1 \right)
\left( \frac{t}{t_s} \right)^{2\gamma+1} \right]^3}
\left( \frac{1}{t_s} \right)
\left( \frac{t}{t_s} \right)^{2\gamma}
\left[ 1 - \left( \frac{t}{t_s} \right)^{2\gamma+1}
\right]\,.
\label{eq:I1-2}
\end{eqnarray}
Eq.~(\ref{eq:I1-2}) implies that the relation $d U(t)/\left(dt\right) <0$ 
is always satisfied because we consider only the period $0<t<t_s$ as 
mentioned above. 
This means that $U(t)$ decreases monotonously. Thus, the value of $U(t)$
evolves from positive to negative. From Eq.~(\ref{eq:I0-1-1}), we see that
the value of $w_\mathrm{eff}$ crosses $-1$.
Once the universe enters the phantom phase, it stays in this phase, 
namely, the value of $w_\mathrm{eff}$ remains less than $-1$, and
finally the Big Rip singularity appears because $U(t)$ decreases 
monotonically. 
We note that there could be other types of the finite-time future 
singularities in modified gravity as shown in 
Refs.~\refcite{B-N-O,N-O-PRD78-046006-08}.

By using Eqs.~(\ref{PDF2}), (\ref{PDF4}), (\ref{PDF6}) and (\ref{PDF8}), 
$P(t)$ is obtained as 
\begin{eqnarray}
P(t) = \left[ \frac{\left( \frac{t}{t_0}  \right)^\gamma}
{1-\left( \frac{t}{t_s}  \right)^{2\gamma+1}} \right]^5
\sum_{j=\pm} \tilde{p}_j t^{\beta_j}\,.
\label{eq:I2}
\end{eqnarray}
It follows from Eqs.~(\ref{eq:2.12}) and (\ref{eq:I2}) that 
$Q(t)$ is given by 
\begin{eqnarray}
Q(t) = -6H
\left[ \frac{\left( \frac{t}{t_0} \right)^\gamma}
{1-\left( \frac{t}{t_s}  \right)^{2\gamma+1}} \right]^5
\sum_{j=\pm} \left( \frac{3}{2}H + \frac{\beta_j}{t} \right)
\tilde{p}_j t^{\beta_j}\,.
\label{eq:I3}
\end{eqnarray}
If Eq.~(\ref{eq:I0-2}) can be solved with respect to $t$ as $t=t(R)$, 
in principle we can find the form of $F(R)$ by using this solution and 
Eqs.~(\ref{eq:2.4}), (\ref{eq:I2}) and (\ref{eq:I3}). 
However, for the general case it is difficult to solve 
Eq.~(\ref{eq:I0-2}) as $t=t(R)$. Therefore, as an solvable example, we 
illustrate the behavior of $t_s^2 F(\tilde{R})$ as a function of 
$\tilde{R} \equiv t_s^2 R$ in Fig.~1 for $\gamma =1/2$, 
$\tilde{p}_+ =-1/t_s^{\beta_+}$, $\tilde{p}_- =0$, 
$\beta_+ = \left(1+2\sqrt{19}\right)/2$ and $t_s =2t_0$. 
The quantities in Fig.~1 are described in dimensionless quantities. 
The horizontal and vertical axes show $\tilde{R}$ and $t_s^2F$, respectively. 
(Here, $\tilde{R} = t_s^2R = 4R/R_0$, where $R_0$ is the current curvature. 
In deriving this relation, we have used $t_s =2t_0$, $t_0 \approx H_0^{-1}$, 
where $H_0$ is the present Hubble parameter.) From Fig.~1, we see that 
the value of $F(R)$ increases as that of $R$ becomes larger.

\begin{figure}[tbp]
\begin{center}
   \includegraphics{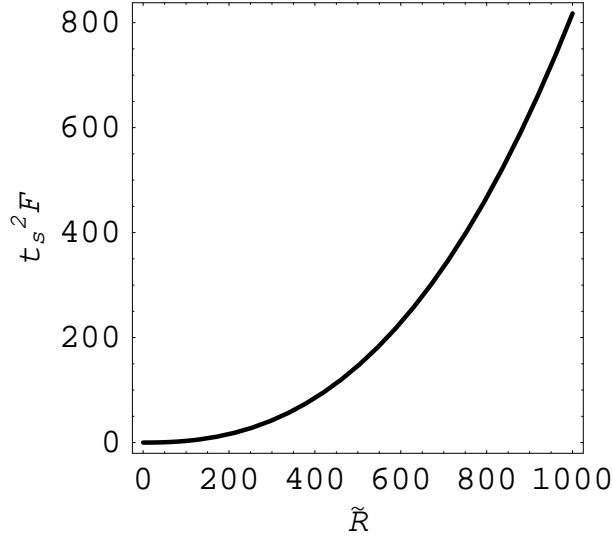}
\caption{Behavior of $t_s^2F(\tilde{R})$ as a function of $\tilde{R}$ for 
$\gamma =1/2$, $\tilde{p}_+ =-1/t_s^{\beta_+}$, 
$\tilde{p}_- =0$, $\beta_+ = \left(1+2\sqrt{19}\right)/2$ and $t_s =2t_0$. 
}
\end{center}
\label{fg:1}
\end{figure}

To explore the analytic form of $F(R)$ for the general case, we examine 
the behavior of $F(R)$ in the limits $t \to 0$ and $t \to t_s$. 
When $t \to 0$, from Eq.~(\ref{PDF10}) we get 
\begin{eqnarray}
t \sim \sqrt{ \frac{60\gamma \left( 20\gamma -1 \right)}{R} }\,.
\label{eq:I4}
\end{eqnarray}
In this limit, it follows from Eqs.~(\ref{eq:2.4}), (\ref{PDF10}),
(\ref{eq:I2}), (\ref{eq:I3}) and (\ref{eq:I4}) that 
the form of $F(R)$ is given by 
\begin{eqnarray}
F(R) \Eqn{\sim}
\left\{
\frac{\left[\frac{1}{t_0} \sqrt{60\gamma \left( 20\gamma -1 \right)} R^{-1/2}
\right]^\gamma}{1 -
\left[\frac{1}{t_s} \sqrt{60\gamma \left( 20\gamma -1 \right)} R^{-1/2}
\right]^{2\gamma+1}} \right\}^5 R \nonumber \\
&& \hspace{10mm}
{}\times
\sum_{j=\pm}
\biggl\{ \left( \frac{5\gamma -1 -\beta_j}{20\gamma -1} \right) \tilde{p}_j
\left[60\gamma \left( 20\gamma -1 \right) \right]^{\beta_j /2}
R^{-\beta_j /2}
\biggr\}\,.
\label{eq:I5}
\end{eqnarray}

Note that such action belongs to general class of actions with positive and 
negative powers of curvature introduced 
in Ref.~\refcite{Nojiri:2003ft}. 

On the other hand, when $t\to t_s$, from Eq.~(\ref{PDF12}) we obtain
\begin{eqnarray}
t \sim t_s - 3\sqrt{ \frac{140}{R} }\,.
\label{eq:I6}
\end{eqnarray}
In this limit, it follows from Eqs.~(\ref{eq:2.4}), (\ref{PDF12}),
(\ref{eq:I2}), (\ref{eq:I3}) and (\ref{eq:I6}) that
the form of $F(R)$ is given by
\begin{eqnarray}
F(R) \Eqn{\sim}
\left(
\frac{
\left\{ \frac{1}{t_0}
\left[ t_s - 3\sqrt{140} R^{-1/2} \right]
\right\}^\gamma}
{1 -
\left[ 1 - \frac{3\sqrt{140}}{t_s} R^{-1/2}
\right]^{2\gamma+1}
} \right)^5 R
\sum_{j=\pm}
\tilde{p}_j
\left[ t_s - 3\sqrt{140} R^{-1/2}
\right]^{\beta_j}
\nonumber \\
&& \hspace{0mm}
{}\times
\Biggl\{
1- \sqrt{\frac{20}{7}}
\left[
\sqrt{\frac{15}{84}} t_s
+ \left( \beta_j - 15 \right) R^{-1/2}
\right]
\frac{1}{t_s - 3\sqrt{140} R^{-1/2}}
\Biggr\}\,.
\label{eq:I7}
\end{eqnarray}
For large $R$, namely, $t_s^2R \gg 1$, the expression of $F(R)$ in 
(\ref{eq:I7}) can be approximately written as 
\begin{eqnarray}
F(R) \approx
\frac{2}{7}
\left[
\frac{1}{3\sqrt{140} \left( 2\gamma +1 \right)}
\left( \frac{t_s}{t_0} \right)^\gamma \right]^5
\left(
\sum_{j=\pm}
\tilde{p}_j t_s^{\beta_j} \right)
t_s^5 R^{7/2}\,.
\label{eq:I8}
\end{eqnarray}

\section{Summary}

We have studied a crossing of the phantom divide in $F(R)$ gravity. 
We have reconstructed an explicit model of $F(R)$ gravity in which 
a crossing of the phantom divide can occur 
by using the reconstruction method.\cite{Nojiri:2006gh,Nojiri:2006be} 
As a result, we have shown that the Big Rip singularity may appear 
in the reconstructed model of $F(R)$ gravity. 
We finally mention that by adding $R^2$ term (as it was first proposed in 
Ref.~\refcite{abdalla}) to the model or 
by adding non-singular theory,\cite{Nojiri:2009xw} 
$R^2 \left(R^n+c_1\right)/\left(R^n+c_2\right)$, 
where $n$, $c_1$ and $c_2$ are constants, Big Rip singularities could be 
avoided because phantom behavior becomes 
transient.\cite{B-N-O,N-O-PRD78-046006-08}

\section*{Acknowledgments}

The work by K.B. and C.Q.G. is supported in part by
the National Science Council of R.O.C. under:
Grant \#s: NSC-95-2112-M-007-059-MY3 and
National Tsing Hua University under Grant \#:
97N2309F1 (NTHU),
that
by S.D.O. was
supported in part by MEC (Spain) projects FIS2006-02842 and
PIE2007-50I023, RFBR grant 06-01-00609 and LRSS project N.2553.2008.2,
and
that by S.N. is supportedby Global
COE Program of Nagoya University provided by the Japan Society
for the Promotion of Science (G07).

%
%

\end{document}